\documentclass[letterpaper,twocolumn,10pt]{article}
\usepackage{usenix}
\usepackage{authblk}

\usepackage{graphicx}
\usepackage[normalem]{ulem}
\usepackage{comment}
\usepackage{subcaption}
\usepackage{algpseudocode} 
\usepackage{graphicx}
\usepackage{subcaption}

\usepackage{tikz}
\usepackage{graphicx}
\usepackage{tabularx}
\usepackage{color}
\usepackage{booktabs}
\usepackage{bbding}
\usepackage{pifont}
\usepackage{amssymb}
\usepackage{xtab}
\usepackage{paralist}

\usepackage{svg}
\usepackage{verbatim} 
\usepackage{listings}

\usepackage{xcolor}
\usepackage{xspace}
\newif\ifcommentary
\commentaryfalse

\ifcommentary
\newcommand{\todo}[1]{{\color{red}\textbf{TODO:} #1}\xspace}
\newcommand{\fixme}[1]{{\color{red}\textbf{!! FIXME: #1 !!}}}
\newcommand{\msm}[1]{[{\color{magenta}msm: #1}]}
\newcommand{\Eman}[1]{[{\color{blue}Eman: #1}]}
\newcommand{\santiago}[1]{[{\color{green}Santiago: #1}]}
\newcommand{\chinenye}[1]{[{\color{orange}Chinenye: #1}]}

\else
\newcommand{\todo}[1]{}
\newcommand{\fixme}[1]{}
\newcommand{\msm}[1]{}
\newcommand{\Eman}[1]{}
\newcommand{\santiago}[1]{}
\newcommand{\chinenye}[1]{}
\fi
\newcommand{\system}{Petra\xspace}

\newcommand{\eg}{e.g.,\xspace}
\newcommand{\ie}{i.e.,\xspace}

\newcommand{\Paragraph}[1]{\noindent\textbf{#1}}

\usepackage{amsmath}

\usepackage{filecontents}
\usepackage[ruled, linesnumbered]{algorithm2e}
\usepackage{etoolbox}
\usepackage[available]{usenixbadges}

\makeatletter
\patchcmd\UrlBreaks{\do\@}{\do\@\do\-}{}{}
\makeatother

\begin{document}
\pagestyle{empty}

\date{}

\title{\Large \bf Trustworthy and Confidential SBOM Exchange}

\author{
{\rmfamily
Eman Abu Ishgair$^{\dagger}$\quad
Chinenye Okafor$^{\dagger}$\quad
Marcela S. Melara$^{\ddagger}$\quad
Santiago Torres-Arias$^{\dagger}$}\\
{\small $^{\dagger}$Purdue University, \texttt{\{eabuishg, okafor1, santiagotorres\}@purdue.edu}}\\
{\small $^{\ddagger}$Intel Corporation, \texttt{marcela.melara@intel.com}}
}

\maketitle

\begin{abstract}
Software Bills of Materials (SBOMs) have become a regulatory requirement for improving software supply chain security and trust by means of transparency regarding components that make up software artifacts.
However, enterprise and regulated software vendors commonly wish to restrict who can view confidential software metadata recorded in their SBOMs due to intellectual property or security vulnerability information. 
To address this tension between transparency and confidentiality, we propose \system, an SBOM exchange system that empowers software vendors to interoperably compose and distribute redacted SBOM data using selective encryption.
\system enables software consumers to search redacted SBOMs for answers to specific security questions
without revealing information they are not authorized to access.
\system leverages a format-agnostic, tamper-evident SBOM representation to generate efficient and confidentiality-preserving integrity proofs, allowing interested parties to cryptographically audit and establish trust in redacted SBOMs. 
Exchanging redacted SBOMs in our~\system prototype requires less than 1 extra KB per SBOM, and SBOM decryption accounts for at most 1\% of the performance overhead during an SBOM query.
\end{abstract}

\section{Introduction}
\label{sec:introduction}
Millions of software vendors today build products that depend on open and closed source software components distributed through a myriad of channels; each component is, in turn, composed of various software dependencies.
While open source software is typically available for inspection, accurately identifying the direct and transitive dependencies that are redistributed in a software product or a deployed system remains a difficult challenge~\cite{stalnaker_boms_2024,dalia_sbom_2024}.

Motivated by large-scale software supply chain incidents such as the Log4Shell vulnerability~\cite{log4j_vuln1} and the SolarWinds hack~\cite{solarwinds-report}, interest in Software Bill of Materials (SBOM) has significantly increased for their ability to provide detailed documentation about the components in a software artifact.
Recent regulations~\cite{executive_order, EU_CRA} cemented the need for generating SBOMs to enhance the transparency of the software supply chain, prompting software vendors to collect and disseminate SBOMs for their software products.

Yet, software vendors face three key challenges that hamper the adoption of SBOMs in their software development lifecycles.
\textbf{1) Data confidentiality:} Enterprise and government practitioners have highlighted significant concerns about  the disclosure of intellectual property or private data when sharing SBOMs with software consumers~\cite{xia_sbom_study_2023,bi_way_2024,zahan_software_2023,Software_Consumers_Playbook,Software_Suppliers_Playbook,stalnaker_boms_2024}.
\textbf{2) Trustworthy SBOM data:} Another common problem is the lack of standard mechanisms to verify the integrity of SBOM data and establish its trustworthiness~\cite{xia_sbom_study_2023,dalia_sbom_2024, stalnaker_boms_2024}.
\textbf{3) Interoperable exchange:} Additionally, current tools are largely unable to generate SBOMs that can be \emph{composed and (re)distributed} in a standardized and efficient manner that enables cross-vendor interoperability~\cite{xia_sbom_study_2023,kloeg_charting_2024,zahan_software_2023}.

Current approaches for secure and trustworthy SBOM exchange rely on contractual/legal processes like licensing and non-disclosure agreements~\cite{Software_Suppliers_Playbook}, or disparate ad-hoc systems~\cite{dalia_sbom_2024,xia_sbom_study_2023,lf-sbom-readiness-report} (\eg secure repositories with restricted access~\cite{esf-report} or dedicated publish-subscribe systems\cite{ntia-sbom-sharing}).
While these approaches facilitate the sharing of a vendor's SBOM, they offer limited scalability and interoperability for secure SBOM exchange because they focus narrowly on distributing SBOMs to \emph{direct} software consumers.
Thus, as SBOM data completeness is varied~\cite{stalnaker_boms_2024,zahan_software_2023}, consumers further downstream the supply chain still face challenges discovering and trusting SBOMs for transitive dependencies.

To address these challenges, we propose \system, an interoperable system for confidential SBOM exchange and trust.
Petra is the first-of-its-kind SBOM exchange system to support interoperable composition and redistribution of SBOMs while enforcing confidentiality and enabling cryptographic auditing of redactions. Unlike existing SBOM exchange mechanisms, Petra preserves integrity and access control after SBOMs are redistributed across organizational boundaries.



\Paragraph{SBOM exchange architecture for interoperability.}
At the foundation of \system's design is an abstract SBOM exchange architecture that centers the ubiquitous practice of software composition and redistribution, 
\ie the propagation of software components incorporated into other software products.
Our model extends this paradigm to SBOMs, enabling the generation and exchange of \emph{composable} SBOMs in a manner agnostic to SBOM data format or software ecosystem.

\system achieves this goal at two levels.
At the system level, \system defines a set of abstract SBOM exchange roles and protocols that can be instantiated in existing software supply chains (\S\ref{sec:system-model}).
At the data level, \system breaks down SBOM information into an \emph{SBOM tree}, a tree-based data structure that mirrors the documented relationships between software components and their metadata (\S\ref{secsec:sbom-trees}).

For instance, a software vendor \textit{Acme} builds a software package \texttt{foo} that depends on package \texttt{bar}. With \system, \textit{Acme} will generate an SBOM tree for package \texttt{foo} 
that at a high level will have \texttt{foo} at its root node, and a child node for \texttt{bar}. 

\system SBOM trees enable efficient searching and \emph{merging} of SBOM information while ensuring that software consumers can trace a particular dependency back to its origin.
This allows \system to support SBOMs that reflect the highly compositional nature of today's software and retain transitive dependency information about an artifact within a single SBOM, obviating the need to distribute per-dependency SBOMs.

\Paragraph{Confidentiality with selective redaction.}
SBOM trees allow \system to balance the tension between transparency and confidentiality in a customizable way. 
We retain the basic model of SBOM generators creating SBOMs for software products, but enable vendors to control access to their own SBOM information through selective redaction, even if this information is contained within another vendor's SBOM.

Specifically, leveraging attribute-based encryption,
\system is capable of redacting information at multiple levels of the SBOM tree (\S\ref{secsec:cp-abe}).
\system then captures organization-level requirements via namespace attributes (\eg company domains).

\Paragraph{Cryptographically establishing trust in SBOM data.}
While placing an entire SBOM within a single Merkle tree node (analogous to binary transparency~\cite{binary-transparency}, for example) would enable detection of unexpectedly issued SBOMs, such a structure would still need ad-hoc mechanisms for data lookups within SBOMs, SBOM composition, and auditable SBOM re-distribution.
Thus, to detect modifications~\cite{xia_sbom_study_2023,bi_way_2024} of the contents of an SBOM, even when redacted, \system computes an \emph{authenticated Merkle tree} on top of an SBOM tree (\S\ref{secsec:merkle-trees}).

c
While verifying the integrity of a Merkle SBOM tree requires reconstructing the entire structure to validate its root, users may also delegate this operation to trusted verifiers, not unlike monitors in Certificate Transparency~\cite{laurie2014certificate}.
On the other hand, the Merkle SBOM tree also allows \system to efficiently prove \emph{sameness} between redacted and plaintext SBOM data while maintaining the confidentiality of the remainder of the tree's contents.

Checking stronger \emph{completeness} and \emph{accuracy} properties of SBOM information is considerably harder to achieve in an interoperable, data-agnostic manner as this requires cross-referencing the contents of an SBOM with various application-aware data sources~\cite{bi_way_2024}.
Thus, \system establishes cryptographic integrity as the foundation for trustworthy SBOMs.

\Paragraph{Petra implementation.}
We instantiate \system in a prototype system, which includes an ecosystem-agnostic redacted SBOM generator, as well as a client-side application for SBOM consumers and verifiers.
The generator implements \system's SBOM tree, selective redaction based on CP-ABE~\cite{bethencourt-cpabe-2007} and sameness proof generation, while the client-side tool supports queries on redacted SBOMs and proof verification. 
A separate verifier component showcases \system Merkle SBOM tree integrity checking. 
A Fulcio-backed~\cite{sigstore_fulcio, newman_sigstore_2022} key management server handles signing and redaction keys for generators and consumers. 
We study the storage overhead of our evaluation and find that it adds acceptable storage overhead --- 13\% compared to their non-encrypted counterparts.
Similarly, encryption, decryption, and membership routines add less than one second  overhead for average SBOM sizes, and two seconds at a maximum.



\section{Related Work}
\label{sec:related work}


\subsection{Software Supply Chain Security \& SBOMs}
The last half-decade has evidenced a widespread exploration of supply chain security techniques from industry, academia and government.
Various efforts aim to provide security enhancements through transparency, integrity, and separation~\cite{okafor_sok_2022}.
Frameworks such as in-toto~\cite{torres-arias_-toto_2019} provide end-to-end integrity auditing across phases of the software 
development lifecycle. Other frameworks like The Update Framework (TUF)~\cite{kuppusamy-diplomat-2016}, 
CHAINIAC~\cite{nikitin2017chainiac}, and Sigstore~\cite{newman_sigstore_2022} focus on ensuring the integrity of distributed software.

Binary transparency\cite{al-bassam_contour_2018} provides cryptographic commitments to software binaries on a public, 
append-only log, making it easier to detect misbehavior. In contrast, Petra aims to balance transparency and confidentiality of software through selective redaction and cryptographic integrity.

Software Bills of Materials (SBOMs) have been recognized as a critical aspect of software supply chain security as they provide 
visibility into the components of a software product. SBOMs are meant to enable informed decision-making on security 
risks, license compliance, and overall software quality. Government organizations have developed standards 
for SBOMs, outlining minimum elements for dependency tracking~\cite{ntia2021} and defining various SBOM types~\cite{cisa2023}
to address varying needs at different stages of the software lifecycle.

Although SBOMs are now recognized as crucial for software supply chain security, their wide-scale adoption in practice
remains limited. A 2022 study by the Linux Foundation revealed that only 18\% of organizations use SBOMs extensively across their 
operations~\cite{lf-sbom-readiness-report}. A separate study focusing on open source projects on GitHub, 
where only 186 projects utilizing SBOM generation tools were found~\cite{nocera_software_2023}.

This limited adoption has been attributed to social factors like unclear benefits and use 
cases~\cite{chaora2023discourse,melara_translation_2023} and a lack of awareness about SBOMs~\cite{nocera_software_2023}, concerns about sensitive information disclosure~\cite{xia_sbom_study_2023,dalia_sbom_2024,zahan_software_2023} and 
the lack of effective SBOM analysis tooling~\cite{xia_sbom_study_2023, stalnaker_boms_2024}.
Our approach in \system specifically seeks to address unwanted information disclosure.

\Paragraph{Interoperable SBOM exchange.}
Compatibility issues between SBOM generators and consumers arise because tooling has inconsistent support for 
the two common SBOM data exchange formats SPDX~\cite{noauthor_spdx_nodate} and CycloneDX~\cite{noauthor_cyclonedx_nodate}.
To address these issues, open source initiatives such as the Open Source Security Foundation's ProtoBOM~\cite{protobom} and bomctl~\cite{bomctl} projects seek to provide mappings between fields on either standard to allow for translation.

Although limited options for standardized SBOM exchange platforms hinder interoperability across tools and 
vendors~\cite{bi_way_2024, xia_sbom_study_2023}, efficient SBOM and supply chain metadata exchange remains a relatively 
understudied area.
Frameworks like the CycloneDX SBOM Exchange API~\cite{cyclonedx_transparency_exchange_api_2024}, Federated PURL (package 
URL)~\cite{purl_spec_2024}, DBOM~\cite{dbom} and GUAC~\cite{guac} exist to facilitate 
metadata exchange, but there is a lack of uniformity and established sharing mechanisms\cite{xia_trust_2024, dalia_sbom_2024} and 
academic literature exploring this problem remains limited.
Approaches such as Manufacturer Usage Descriptions (MUD) for IoT devices~\cite{mud} exist to share information 
with consumers, but have limited broad deployments.

Petra builds upon existing efforts on interoperable tools that handle SBOM data exchange~\cite{kloeg_charting_2024,xia_sbom_study_2023,zahan_software_2023}. 
However, existing SBOM exchange frameworks fundamentally differ from Petra in the security guarantees they provide. Systems such as the CycloneDX SBOM Exchange API, GUAC, DBOM, and related metadata exchange platforms primarily address interoperability and metadata exchange, but offer little support for confidentiality, auditable redaction, or composition of SBOMs originating from different vendors. Relying on repository access controls or API authentication, these alternatives to \system cannot guarantee integrity once SBOMs are redistributed. Petra addresses this gap by binding selective disclosure and integrity guarantees directly to the SBOM structure itself, enabling confidentiality- and integrity-preserving SBOM composition and redistribution without relying on trusted repositories or centralized enforcement. 

Much like ProtoBOM, Petra is format-agnostic and uses a graph-based SBOM representation and transformations, supporting both SPDX and CycloneDX formats. The OWASP Transparency Exchange API (TEA)~\cite{cyclonedx_transparency_exchange_api_2024} provides SBOM data "on the fly," effectively acting as a generator for a particular stream. In practice, \system complements the current tools, and we have engaged with the ProtoBOM developers to integrate its redaction features in bomctl~\cite{bomctlroadmapmd_nodate}.

\Paragraph{SBOM quality \& accuracy.}
Studies have raised concerns about the quality and accuracy of SBOMs due to incomplete data\cite{kloeg_charting_2024}, inconsistent 
results from different tools~\cite{o2024impacts}, and a lack of validation mechanisms\cite{kloeg_charting_2024}. These inconsistencies 
were attributed to reliance on heuristic methods, incomplete package manager metadata, and difficulties in handling statically linked 
binaries during SBOM generation~\cite{zahan_software_2023} compounded with unclear definitions of SBOM
quality~\cite{torres_sbom-quality_2023}. While \system does not aim to address these issues, we focus on integrity and confidentiality as foundational preconditions for quality, 
because SBOM quality and accuracy cannot be meaningfully established without trust in the authenticity and protection of SBOMs. 

\subsection{Privacy-preserving Data Exchange} 
Privacy-enhancing techniques have been a central topic in cryptography for many decades.
Various approaches to secretly share private information have been widely explored in fields such as Private Information Retrieval 
(PIR)~\cite{yekhanin2010private}, as well as Homomorphic Encryption (HE)~\cite{chotard2016homomorphic} and Secure Multiparty Computation (SMC) ~\cite{bayatbabolghani2018secure}.
There are applications of these in the area of supply chain security~\cite{cappos2013avoiding, gittuf_repo}.
Similarly, other supply chain security systems that further privacy are Speranza~\cite{merrill_speranza_2023} and Gringotts~\cite{xu2023gringotts}, which protect software distribution security and privacy and version control system security and privacy respectively.

As for non-cryptographic approaches, a promising alternative 
is Trusted Execution Environments (TEE~\cite{jauernig2020trusted, widevine}), which can allow to encrypt SBOMs wholesale and provide access to users using a messaging interface. 
Perhaps the most important distinction is that attribute-based encryption (ABE) allows \system to not require a live service to re-distribute SBOM information, and instead leans on key management to issue keys to parties in the ecosystem. 

\Paragraph{Applications of CP-ABE}
Several works have applied Ciphertext-Policy Attribute-Based Encryption (CP-ABE) in other domains like cloud data sharing ~\cite{li_phr_2013, hur_revocation_2011, xu_lightweight_2018}, electronic health records~\cite{etsi2018ts103532}, and IoT/edge computing~\cite{han2018efficient,zhang_healthcareiot_2020}. These systems highlight the applicability of CP-ABE in practical settings in real world environments.  In the SSC domain,
a concurrent work~\cite{xia_trust_2024} proposes a blockchain-based SBOM exchange architecture that uses verifiable credentials to enable verification 
of SBOM generation, and ABE for selective SBOM data disclosure. More specifically,~\cite{xia_trust_2024} introduces an 
oversight authority to determine generators' adherence to SBOM regulations, and rely on various on- and off-chain components 
to coordinate SBOM generation and distribution.

While Petra shares the goal of selectively disclosing SBOM data, our Merkle SBOM tree-based design is meant to integrate and 
interoperate with existing tools and software supply chain components, rather than requiring dedicated and costly blockchain-based infrastructure to provide integrity checking. 
As such, \system can offer software producers additional flexibility in defining SBOM access control policies.



\section{System and Threat Model}
\label{sec:system-model}
\system aims to provide a SBOM exchange system that facilitates
trustworthy and interoperable SBOM generation and distribution
that meets the diverse confidentiality needs of today's software vendors.


\Paragraph{SBOM Definition.} Per the U.S. NTIA~\cite{ntia_sbom}, we define a Software Bill of Materials (SBOM)
as a record that unambiguously describes a software artifact's components, their relationships, and relevant metadata, such as vulnerability information, each dependency's license, and version.
Thus, the baseline data fields used to describe a software artifact in an SBOM include supplier name, component name, component version, unique identifiers, dependency relationships and SBOM author~\cite{ntia_sbom}. In Petra, we assume that SBOMs follow a given schema which specifies the \emph{fields} that contain the different elements of an SBOM.
The common schemas used to generate and consume SBOMs are SPDX~\cite{noauthor_spdx_nodate}, CycloneDX~\cite{noauthor_cyclonedx_nodate}, and SWID tags~\cite{swid-tags}.
We include a snippet of an SPDX SBOM in Appendix~\ref{appendix: SBOM example}.

\begin{figure}[t]
    \centering
    \includegraphics[width=\columnwidth]{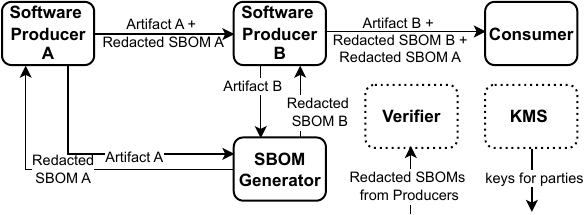}
    \caption{ Abstract SBOM exchange system. A generator creates redacted SBOMs according to producers policies', which are then redistributed along artifacts to other producers or consumers through the redistribution chain. We omitted distributors for simplicity.}
    \label{fig:Private-Arch}
\end{figure}

\subsection{SBOM Exchange Roles and Processes}
\label{secsec:roles-processes}

As depicted in Fig.~\ref{fig:Private-Arch}, we propose an abstract SBOM exchange system involving the following roles and processes:

\Paragraph{Software Producer.} An important prerequisite for SBOM exchange is the creation of software artifacts by software producers. 
In practice, a software producer may be an individual engineer on an open source project, or a larger
organization such as an enterprise or government entity. 

Generally, the software producer aims to release the software
artifact publicly, to a select set of recipients, or internally within an organization. 
Thus, depending on the publication scenario, the software producer may seek to restrict access to the software artifact as a whole, or to (portions of) the software's metadata, including its SBOM. 
In confidential SBOM exchange, the producer is responsible for authoring the redaction policy that governs which parts of the SBOM are visible to whom. 
This policy encodes confidentiality requirements based on vendor preferences or regulatory obligations. 

\Paragraph{Generator.} Following the creation of a software artifact, the SBOM generator creates an SBOM that lists the components
and related metadata of a given software artifact on behalf of its software producer.
In a confidential SBOM exchange system, the generator also applies and preserves the producer's SBOM data access policy requirements, even if the SBOM is composed with another and/or redistributed at a later time.

The separation of generator and producer is deliberate: Petra aims to accommodate two different SBOM generation use cases. 
First, there are practitioners for whom a single SBOM generator is not sufficient to produce a comprehensive and composed SBOM for their software, as evidenced by the CISA SBOM types ~\cite{cisa2023}, where multiple generators record different elements of an SBOM document. 
Second, and more crucially, is the plethora of third-party generator tools and services such as FOSSA, GitHub, and others (\eg~\cite{gh-attest-sbom-action,gh-export-sbom,amazon-sbom-inspector,blue-goat-sbom-services,soos-research-packages}). 
Thus, the explicit generator role of Petra offers a flexible architecture, but this design choice does not preclude merging of the two roles in specific implementations. 


\Paragraph{Distributor.} Once an SBOM is generated, the software producer or generator submits the SBOM to a distributor making it available for discovery. 
To this end, the distributor provides a system for SBOM storage and management, which commonly allows for retrieving an SBOM for a target artifact, e.g., via a global identifier such as a pURL~\cite{purl_spec_2024}, a standardized identifier format used across major SBOM standards (e.g., pkg:npm/express@4.17.1). 
Such SBOM databases or hubs (\eg~\cite{rkvst, endor-labs-sbom-hub, lineaje-sbom-hub}) may also support analysis queries on stored SBOM information without delivering a full SBOM document.

Alternatively, SBOM distribution occurs when an SBOM is delivered with the software artifact (i.e., in-band distribution). 
For instance, the distributor is hosted on the same platform  that delivers software artifacts, such as social coding platforms~\cite{github} and package repositories~\cite{npm}.
Similarly, if a software producer distributes an SBOM along with their software artifact, they act as a distributor.

\Paragraph{Consumers.} Individual software developers and organizations consume software artifacts and their associated SBOMs, 
either for direct deployment or to integrate an artifact as a dependency and redistribute it in their own software product. 
In both cases, consumers leverage SBOMs to gain security insights about consumed software by interacting with distributors to obtain the needed SBOM documents.

When acting as a software producer, the consumer creates a software artifact that is composed of various dependencies, resulting in a \emph{composed SBOM} from a generator. 
That is, say the consumer's software package \texttt{foo} includes dependency package \texttt{bar}, the SBOM $S_{foo}$ should contain SBOM $S_{bar}$.

\Paragraph{Verifier.} To ensure that SBOM information is trustworthy, verifiers obtain SBOMs from distributors for the purpose of validating their format and contents. 
Specifically, verifiers check that a distributed SBOM was not tampered with and matches the output of a generator, as well as checking whether an SBOM reflects a software artifact's composition.

While other roles may act as verifiers, we envision trusted third-party verifiers will commonly perform this task on their behalf. 
For example, a government agency or dedicated auditor organization may act as a verifier checking individual SBOMs or portions of the software supply chain.

\Paragraph{Key Management Service (KMS).}
To reduce the burden of key management for other participants, the KMS acts as a trusted central authority for key generation, user onboarding, key distribution, revocation, and expiration. 

\subsection{Threat Model}
\label{subsec: threat-model}
Confidential SBOM exchange is concerned with the following threats,
labeled as \textbf{(T1–T7)} for reference in \S~\ref{sec:security-analysis}. 

Software producers and generators are untrusted as they may be driven by business incentives to circumvent security or compliance audits for targeted consumers.
That is, software producers and generators may \textbf{tamper with SBOM data (T1)}~\cite{ozkan2025supplychaininsecuritylack}, either by omitting certain fields such as vulnerabilities, or altering fields such as dependency versions.
Similarly, distributors are untrusted in confidential SBOM exchange as they might tamper with stored SBOM data to undermine trust in the software supply chain of a major software vendor, 
or to give false sense of security regarding a software artifact.

Distributors and generators may also seek to mount a \textbf{split view attack (T2)}~\cite{melara_coniks_2015} in an attempt to fool targeted consumers into using a compromised artifact.
Specifically, generators may create inconsistent copies of redacted SBOMs for the same software product, while distributors may provide modified or outdated copies of an SBOM to different consumers.  
In the case of SBOM redistribution, malicious distributors or consumers may cause a denial-of-service (DoS) by \textbf{corrupting SBOMs redistributed to downstream consumers (T3)}~\cite{nikitin2017chainiac}. 

Consumers have also been shown to
leverage SBOM information, such as package versions and reported vulnerabilities, for malicious purposes~\cite{dalia_sbom_2024, kloeg_charting_2024, xia_sbom_study_2023, zahan_software_2023}.
Hence, software producers are concerned with \textbf{unauthorized SBOM data leaks (T4)} by untrusted consumers.

Consumers may further seek to circumvent any access restrictions or redaction on SBOM information put in place by generators.
That is, 
consumers with partial access to SBOMs may attempt a \textbf{dictionary attack (T5)}~\cite{pinkas2002securing} on encrypted fields to infer their values.
Additionally, consumers with different access permissions to an SBOM may \textbf{collude to decrypt inaccessible information (T6)}~\cite{bethencourt-cpabe-2007} about a competitor's products in an SBOM,
or collude to share SBOM data they can each access to \textbf{\emph{infer} inaccessible SBOM data (T7)}~\cite{naveed2015inference}.


\Paragraph{Trusted roles and components.}
Verifiers are trusted to faithfully audit redacted SBOMs. Similarly, we assume Key Management Services (KMS) 
faithfully authenticate principals, bind attributes, issue keys, and enforce revocation/expiry. 
KMSes can leverage existing key transparency mechanisms~\cite{newman_sigstore_2022,melara_coniks_2015,malvai_parakeet_2023} for additional cryptographic integrity and compromise detection.
Further, we assume all parties employ cryptographic schemes and compute systems (hardware and OS) that are free of known vulnerabilities. 

\Paragraph{Out of Scope Threats.}
Detecting maliciously crafted SBOMs that conceal malicious components is out of scope, as is detecting malicious components.
These threats are the subject of extensive prior work~\cite{ferreira2021containing,sejfia2022practical,zimmermann2019small,ohm2020backstabber,constantin2018npm}. 
Similarly, detecting flaws with generator tools and processes (e.g., those who may introduce mistakes) is out of scope~\cite{10646983}.
Confidential SBOM exchange also does not attempt to evaluate whether SBOM access policies are excessively restrictive. Detecting or mitigating overly strict 
is a dedicated research area~\cite{xia_trust_2024} and requires separate governance or policy-auditing mechanisms.

\subsection{Design Requirements}
\label{secsec:design-properties}

Derived from our architecture and threat model, a confidential exchange system should meet
the following functional, confidentiality, and integrity requirements. \\

\Paragraph{Functional Requirements.}
\label{subsubsec:functional-properties}


\Paragraph{(F1) Relational SBOM Representation:} SBOMs must be represented using a data structure that preserves the inherent relationships between all software artifacts and associated metadata recorded in an SBOM, regardless of the software ecosystem or data exchange format. 
This supports composition and redaction operations on SBOM data without losing the relational information needed for meaningful software dependency analysis.

\Paragraph{(F2) SBOM Composability:} To reflect software composition, SBOMs must be composable. 
That is, leveraging the relational SBOM representation property, if software artifact $B$ depends on a third-party component $A$, SBOM $S_A$ can be inserted into the contents of SBOM $S_B$.

\Paragraph{(F3) Indexability:} To support interoperable distribution of SBOMs, they must be associated with specific software artifacts via immutable indexes. \\

\Paragraph{Confidentiality Requirements.}
\label{subsubsec:confidentiality -properties}

\Paragraph{(C1) Semantic confidentiality:} To address a core barrier to practical SBOM adoption in enterprise and government settings, 
SBOMs must not leak SBOM contents to parties without explicit permission to access specific pieces of dependency or metadata information. 
This property must hold even during SBOM composition or redistribution. 
In other words, software producers must be able to compose and redistribute SBOM contents without requiring access to any information contained therein.

\Paragraph{(C2) Collusion Resistance:} As consumers may attempt to circumvent semantic confidentiality by colluding to reveal redacted SBOM data,
the selective redaction mechanism must be capable of resisting collusion between parties that have access to only a subset of the needed permissions. \\

\Paragraph{Integrity Requirements.}
\label{subsubsec:integrity-properties}

\Paragraph{(I1) Sameness:} Consumers of a software artifact must be able to verify that redacted SBOM
information faithfully represents and does not omit information from the \emph{plaintext SBOM},
even for redacted data that is only partially accessible. 

Specifically, the original relational SBOM structure must be maintained while ensuring that only the information accessible to a verifier is revealed.
In practice, while access restrictions would allow most consumers to only partially verify sameness, we expect that software producers and select verifiers are permitted to access the full plaintext SBOM and validate full-SBOM sameness.

\Paragraph{(I2) Redistribution Congruence:} To support composability and interoperable distribution of SBOMs, a redistributed SBOM must retain its sameness, and preserve semantic confidentiality as well as collusion resistance. 
That is, an SBOM $S_A$ distributed for software artifact $A$ must remain identical to the SBOM $S_A$ redistributed in SBOM $S_B$ for a larger software artifact $B$ that depends on $A$.

\Paragraph{(I3) Non-equivocation:} Consumers must be able to verify they received the same SBOM as others for a given artifact.
\section{Foundational Data Structures}
\label{sec:Foundational-Techniques}

We instantiate our confidential SBOM exchange architecture in \system.
At its core, \system uses a series of data structures to achieve our design requirements~\S\ref{secsec:design-properties}.
Through a three-stage SBOM process, we use these data structures to transform a textual SBOM document into an auditable, selectively redacted SBOM representation.

\begin{figure*}[t!]

    \begin{subfigure}[t]{0.30\textwidth}
        \centering
       \includegraphics[height=1.7in]{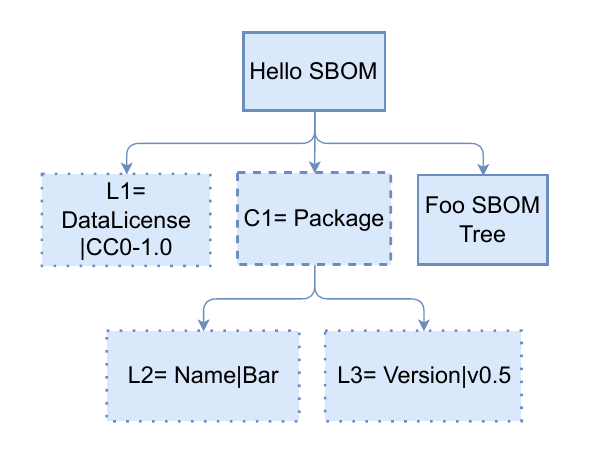}
        \caption{}
        \label{fig:SBOMTree}%

    \end{subfigure}%
~
    \begin{subfigure}[t]{0.34\textwidth}
        \centering
             \includegraphics[height=1.7in]{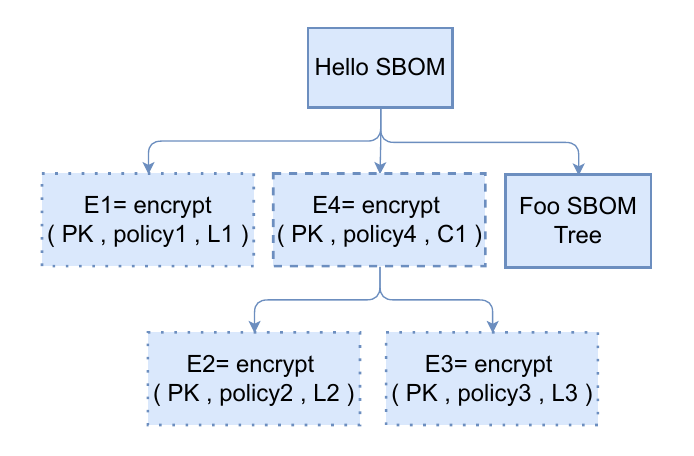}
        \caption{}
                     \label{fig:EncryptedTree}%

    \end{subfigure}
    ~
    \begin{subfigure}[t]{0.32\textwidth}
        \centering
        \includegraphics[height=1.7in]{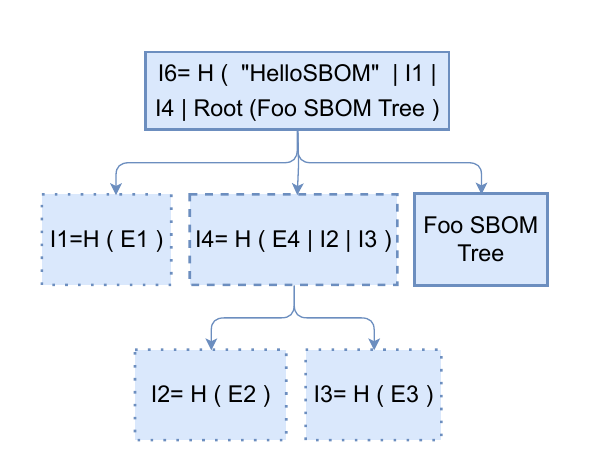}
        \caption{} 
                \label{fig:MerkleTree}%

    \end{subfigure}
    \caption{The Petra SBOM tree is an efficient representation of an (a) SBOM that preserves relational information between dependencies and their metadata.
    This representation also allows Petra to (b) selectively redact and (c) integrity check SBOMs within a single data structure.}
    \label{fig:trees}

\end{figure*}
\subsection{SBOM Tree}
\label{secsec:sbom-trees}

The first stage in \system translates an SBOM document (\eg in SPDX~\cite{noauthor_spdx_nodate} or CycloneDX~\cite{noauthor_cyclonedx_nodate} format)
into a tree data structure that is suitable for efficient redaction and integrity checking.  
We draw upon the insight that an SBOM is a collection of claims about a software artifact.
Namely, each SBOM  describes an aspect of a software product, such as qualitative (e.g., vulnerabilities), legal and compliance (e.g., licenses) and functional (e.g., configurations).

Using this notion, we first \emph{decompose} an SBOM into a series of nodes that describe these claims, and construct an \emph{SBOM tree} that represents the relationships between the elements in the SBOM.
We distinguish two types of tree nodes: 
\emph{Complex nodes} hold \emph{composite} claims, e.g., the collection of metadata about a dependency recorded in an SBOM. 
These nodes allow \system to decompose claims further into a set of \emph{field nodes}, and provide nesting.

Specifically, a field node represents exactly a single claim in an SBOM, such as a license claim, and 
is inserted as leaf nodes into an SBOM tree with the corresponding complex node as its parent in the tree.
For example, a dependency within an SBOM can be decomposed on a series of field nodes, one for each type of metadata
about a dependency such as its name and version number.

At the root of the SBOM tree, \system constructs an \emph{SBOM node} containing a unique indexable identifier for the SBOM to aid in its discovery.
The leftmost tree in Figure ~\ref{fig:trees} (\subref{fig:SBOMTree}) represents an SBOM tree containing another SBOM node.

This construction allows us to meet property \textbf{(F1)}, since the structural properties of the tree describe the inherent semantics of the SBOM.
Further, through this complex node decomposition, Petra can provide property \textbf{(F2)} since a consumer can insert an SBOM within another SBOM by adding it as a complex node within their own SBOM tree.

\subsection{Attribute-Based Redaction}
\label{secsec:cp-abe}

To provide selective redaction of SBOM information, \system utilizes a variant of Attribute-Based Encryption (ABE)\cite{goyal2006attribute} called 
Ciphertext-Policy Attribute-Based Encryption (CP-ABE)\cite{bethencourt-cpabe-2007},
which enables fine-grained access control via selectively encrypted data. 
Unlike traditional public-key encryption, which uses a predefined key for decryption, ABE determines decryption keys based on \emph{attributes} associated with principals and the data itself,
effectively enforcing access control on encrypted data.
These attributes are opaque strings/byte arrays, and encode information about consumers (e.g., role, organization, location). 

The CP-ABE scheme used in \system builds on vanilla ABE schemes by passing the access policy as an encryption parameter. 
The policy is then embedded within the ciphertext, while consumer attributes are embedded in decryption keys. 
This gives software producers the flexibility to define who can access the data based on the attributes of different consumers. 

Thus, \system redaction relies on the following operations:
\begin{enumerate}
    \item Setup(): Generates the public encryption parameter $P_k$ and a master decryption key $M_k$.
    \item Encrypt($P_k,M,A$): Given the public parameter $P_k$, a message $M$, and an access policy $A$, produces a ciphertext.
    \item Key Generation($M_k,S$): Derives a secret decryption key based on a set of consumer attributes $S$ from the master key $M_k$.
    \item Decrypt($P_k,CT,S_k$): Given the public parameters $P_k$, a ciphertext $CT$, and a consumer's secret key $S_k$, produces a plaintext message only if the consumer's attributes satisfy the access policy embedded within the ciphertext.
\end{enumerate}

\Paragraph{CP-ABE on SBOM trees.} 
To provide semantic confidentiality (\textbf{C1}), the second step in \system performs a CP-ABE pass to selectively encrypt individual SBOM tree nodes.
Specifically, we attach an access policy to each node and perform an SBOM tree walk to encrypt the SBOM data in each node, 
namely field names and values, according to the node's policy; multiple nodes may be redacted under the same policy.
We do not encrypt the references to the children of complex nodes, as we still aim to expose the structural relationships of the SBOM.
This allows \system to maintain the functional properties above as well as support SBOM composition.

To reduce the cost of CP-ABE encryption and decryption, we utilize CP-ABE for key encapsulation (CP-ABKEM)~\cite{abkem}. 
Instead of directly encrypting each node with a CP-ABE secret key, we generate a random AES key for each distinct node access policy, and
use the AES key to encrypt all nodes the policy applies to. 
Each AES key is then encrypted using CP-ABE under the policy.
This ensures that only consumers who meet a given policy can decrypt the corresponding AES key and, in turn, decrypt the data. 
Using CP-ABKEM, we leverage the fine-grained access of CP-ABE, and the efficiency of AES. 
Figure~\ref{fig:trees}\subref{fig:EncryptedTree} represents the SBOM tree after encryption.

The choice of CP-ABE is crucial to meet \textbf{(C2)}, since the scheme
requires a decryption key to hold all the attributes needed in order to decrypt a node.
We elaborate further on these security properties in the security analysis (\S\ref{sec:security-analysis}).

\subsection{Merkle SBOM Trees}
\label{secsec:merkle-trees}

The third stage in \system computes cryptographic integrity checks to detect tampering with pre- and post-redaction SBOM trees.
This is done by adapting Merkle tree primitives~\cite{merkle-trees} to our irregularly structured SBOM trees.

A Merkle tree is a tree data structure where each leaf node is referenced by the cryptographic hash of their data block (including children). 
Each non-leaf node, on the other hand, combines the hashes of its child nodes. The tree root, known as the Merkle root hash, represents the hash of all the data in the tree.

Merkle trees can be used to represent authenticated graphs or trees thanks to the fact that each element's identifier is recursively comprised of the hashes of content and children nodes.
This property enables \system to provide indexability \textbf{(F3)} as the Merkle root hash can serve as a content-addressable identifier over an entire (redacted) SBOM~\cite{cisa-rfi-software-identifiers}.

\system achieves sameness \textbf{(I1)} by applying Merkle tree primitives twice. 
First, we compute Merkle hashes over cryptographic commitments (in our design, a hash with a random salt) to the SBOM tree prior to redaction. 
The commitments are used to address the limited keyspace of some SBOM fields and values. 
In a second Merkle pass over an attribute-based redacted SBOM tree, \system captures hashes used to detect tampering with the redacted tree. 


\Paragraph{Membership Proofs.}
\system uses Merkle trees to compute an efficient proof that a particular node is contained within an SBOM tree based on its Merkle root hash. 
    A prover computes an \emph{authentication path}, which consists of the hashes of the sibling nodes along the path from the node in question to the root~\cite{melara_coniks_2015}.
To verify whether a node is a member of the tree, a verifier sequentially recomputes the hashes combined with the sibling hashes, starting from the node in question up to the root,
effectively recreating the Merkle root hash. 
If the computed root hash matches the expected Merkle root, the node must be in the tree.    

Thus, \system uses membership proofs to validate redistribution congruence \textbf{(I2)} without requiring full decryption or reprocessing.
That is, 
during SBOM composition, \system includes the Merkle root of the inserted SBOM tree as a complex node in the parent SBOM tree. 
A membership proof is then generated and embedded into the complex node, linking the inserted redacted SBOM Merkle root to the parent SBOM Merkle structure. 
This allows downstream consumers to verify that the embedded SBOM remains intact and matches its original redacted version \textbf{(I1)}.

\subsection{Building Merkle Redacted SBOM Trees}

In summary, prior to selective redaction, \system computes the Merkle hashes over the commitments to the plaintext data in an SBOM tree using a collision-resistant hash function $H()$.
Then, we construct a selectively redacted SBOM tree by encrypting the SBOM tree nodes using a per-node access policy AES key $K_A$, which is then encapsulated using CP-ABKEM via 
the CP-ABE $Encrypt(P_k,K_A,A)$ operation; for brevity, we use the shorthand $E()$ to represent AES-based node encryption.
In a third pass, \system computes the Merkle hash over the redacted tree node contents.

The construction is as follows:

\Paragraph{Field nodes} containing a field $name$ and field $value$ pair are hashed as:

\begin{center}
    $h_{F_{plain}} = Commit(name||value)$
\end{center}

prior to redaction, and afterwards as:

\begin{center}
    $h_{F} = H(R||A_n||E(salt||name||value)||h_{F_{plain}})$
\end{center}

where $R$ is a constant value used to indicate a redacted node
and $A_n$ represents the node's specific CP-ABE access policy to cryptographically
bind a node's redaction policy to its contents.
$salt$ is the hashed random number used to compute the cryptographic commitment to the plaintext. 

Even though including the access policy $A_n$ results in some information leakage,
an observer may at most \emph{infer} the field name from the attributes allowed
by the access policy, but cannot learn the value of the field.

\Paragraph{Complex nodes} represent an artifact-level SBOM element of type $t$ that has one or more field
nodes $F_0$ to $F_n$ as its children. They are computed as:

\begin{center}
    $h_{C_{plain}} = H(Commit(t)||h_{{F_0}_{plain}}||...h_{{F_n}_{plain}})$
\end{center}

representing the pre-redaction SBOM element relationships, with post-redaction ones captured as:

\begin{center}
    $h_{C} = H(R||A_n||E(salt||t)||h_{C_{plain}}||h_{F_0}||...||h_{F_n})$
\end{center}

Because \system does not aim to hide the structure or relationships between elements in an SBOM to support sameness verification,
we can reveal the complex node's access structure and child nodes without leaking specific field values.

\Paragraph{SBOM nodes} identify an SBOM document and may have field, complex or other SBOM nodes as its children $C_0$ to $C_n$. 
Redaction of SBOM node contents depends on whether the software producer wishes to control access to  SBOM-level data such as its resolvable index or redaction policy.

Regardless, \system hashes the contents of SBOM nodes, redacted or not, to allow for 
integrity checking, authentication, but also indexability of the SBOM. Their hash is:

\begin{center}
    $h_{S_{plain}} = H(Commit(r)||h_{{C_0}_{plain}}||...h_{{C_n}_{plain}})$
\end{center}

where $r$ is the plaintext of any redactable SBOM data. After redaction, an SBOM node
is structured as:

\begin{center}
    $node_{S} = Sig_{K_{s}}(H(Encrypt_{all}(P_k, K_{A_n}, A_n)||h_{S_{plain}}||h_{C_0}||...||h_{C_n}))$
\end{center}

where $K_{s}$ is the generator's SBOM signing key, and $Encrypt_{all}(P_k, K_{A}, A)$ is the CP-ABE encryption of the entire set of per-policy AES encryption keys $K_A$.

\Paragraph{Unredacted nodes.} In practice, we expect SBOM redaction policies that allow for certain SBOM fields to be publicly visible. 
In this case, \system replaces the $R$ redaction indicator with a $P$ constant indicating an unredacted node, computing the Merkle hash over the plaintext field or complex node data directly post-redaction.
\section{\system Processes}
\label{sec:system-processes}
\begin{figure}[t]
    \centering
    \includegraphics[width=\columnwidth]{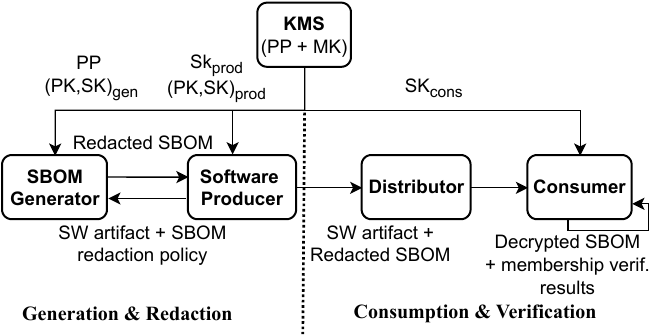}
    \caption{
    In \system, a central KMS generates their redaction and/or signing keys.
    A producer sends the artifact and its redaction policy to a generator, which returns both the pre-redaction plaintext SBOM and the redacted SBOM for cryptographic sameness verification.
    Consumers fetch the artifact and redacted SBOM from distributors, obtain their decryption key from the KMS, and decrypt only authorized fields. 
    }
    \label{fig:Petra-Flow}
\end{figure}

This section presents the following core Petra protocols represented in Figure ~\ref{fig:Petra-Flow}.
\subsection{Redaction Setup}

Before redaction, the KMS generates a series of cryptographic secrets needed for selective encryption/decryption of individual SBOM fields;
these are directly derived from the CP-ABE construction in \S\ref{secsec:cp-abe}. 

\Paragraph{Generators Secrets.}
At startup, the KMS runs CP-ABE setup to generate the public parameters $PP$ used in CP-ABE encryption and decryption, and the master key $(MK_{kms})$ used to derive decryption keys. 
The KMS also generates the generator signing key pair $(SK_{gen}, PK_{gen})$ used to authenticate generated SBOM trees.

\Paragraph{Producers Secrets.}
The KMS generates the producer signing key pair $(SK_{prod}, PK_{prod})$ used to countersign generated SBOM information, in practice, this can be the same signing key used to authenticate/sign SBOM information or attestations 
in existing ecosystems~\cite{in-toto-archivista, sigstore}. 
The KMS makes the public parameter $PP$, the generator's public key $PK_{gen}$, the producer's public key $PK_{prod}$ available to all \system parties.

\Paragraph{Consumer Secrets and Onboarding.}
To access redacted SBOM data, software consumers in \system must obtain decryption keys associated with their access attributes.
To do so, the consumer authenticates herself and her attributes $A$ to the KMS (\eg using 
OIDC~\cite{oidc}).  %
The KMS issues a private decryption key $SK_{A}$ associated with the presented set of attributes $A$, if the consumer can be authenticated. The KMS performs key update process for all non-revoked users, such that decryption keys are updated after $T_{end}$.
\medskip 

\noindent$\mathsf{SK}_{A} \leftarrow \mathsf{CPABE.KeyGen}\big(\mathsf{PP}, \mathsf{MK}_{kms}, A \cup \{\texttt{expiry}: T_{\text{end}}\}\big)$ 
\medskip

\system also makes it possible for the KMS to delegate decryption key issuance to consumers with specific
attributes $D$ (\ie a dedicated secret key $SK_{D}$) to issue a secret sub-key for a subset of their attributes. 
This allows consumers to grant access to a particular subset of attributes (or fields) for members of their organization.

\subsection{Redacted SBOM Generation}
\system extends non-confidential SBOM generation to support selective redaction and SBOM data integrity checking.
That is, \system relies on legacy SBOM generation tools (\eg Trivy~\cite{trivy}) that evaluate a 
particular software supply chain step or artifact to generate an SBOM that 
follows a common format such as SPDX or CycloneDX. 
In this case, a generator receives a set of artifacts and dependency information from the software producer to generate 
a corresponding SBOM.


After generating legacy SBOM data, the \system generator performs a series of passes over it.
As depicted in Algorithm~\ref{alg:redaction}, a software producer submits an $Artifact$, a series of
SBOMs $S$ for dependencies, and a redaction policy $p$ for the artifact.
After computing the redacted SBOM, the generator provides the redacted tree to the producer.
\begin{algorithm}
\footnotesize
\caption{Redaction and Composition}\label{alg:redaction}
\KwData{$Artifact \newline S \leftarrow {S_1, S_2, ..., S_n} \newline p \leftarrow Policy$}
\KwResult{$PlainSBOM, RedactedSBOM$}

    \CommentSty{/* Compose SBOM into tree                        */}
    $PlainSBOM \gets new\;SbomTree()$;
    
    \For{$ S_i \in S$}{
        $st \gets new\;SbomTree(S_i)$
        
        $PlainSBOM.appendChild(st)$
    }
    
    $proof \gets computeSamenessProof(PlainSBOM)$
    
    \CommentSty{/* Apply Encryption using policy $p$ */ }
    $EncSBOM \gets Encrypt(PlainSBOM, p)$

    \CommentSty{/* Merkelize Tree \& Sign */}
    $RedactedSBOM \gets MHT(EncSBOM, proof)$
    
    $Sign(PlainSBOM, SK_{gen})$
    
    $Sign(RedactedSBOM, SK_{gen})$
\end{algorithm}

\begin{figure}[t]
        \centering
        \includegraphics[width=\columnwidth]{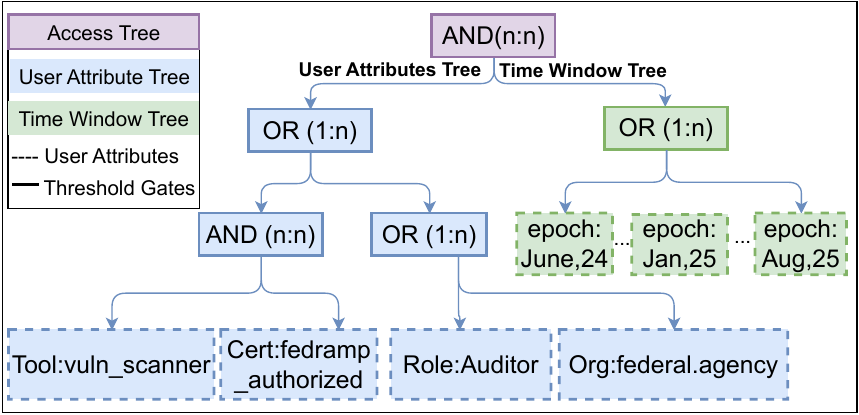}
    \caption{\system Access tree. Two subtrees are used, one for properties tied to identities and roles, and a second one providing a time-bound vector.}
    \label{fig:access-tree}
\end{figure}

\Paragraph{Defining the access policy.} 
Recall that in CP-ABE, permissions are described by an \emph{access tree}, where nodes of the access structure are composed of threshold gates and the leaves describe access attributes where AND gates can be constructed as n-of-n threshold gates and OR gates as 1-of-n threshold gates.
Access attributes are a set of descriptive properties or credentials for a principal, which will be associated with the issued secret key, for example, the user’s role in an organization or project, or an SBOM usage application (vulnerability scanner, license scanner).
Fig.~\ref{fig:access-tree} represents an access tree for the version number field of dependencies, such that version numbers across all dependencies can be accessed by a FEDRAMP~\cite{fedramp} approved vulnerability scanner, an auditor or a federal agency.

In Petra, a policy is a combination of SBOM tree paths and corresponding access trees.
Specifically, a software producer can describe a policy for the tree by describing an access tree policy for a set of fields.
For example, a software producer may allow a \texttt{vulnerability scanner} to access package version numbers in a
redacted SBOM tree, only if the scanner is \texttt{FEDRAMP approved}.
However, a consumer with \texttt{licensing compliance} attributes does not need to know version numbers, but may need to know package names and their corresponding licenses.

\Paragraph{Verifying sameness and counter-signing.}
Once SBOM generation is complete, the software producer ensures that the redacted SBOM 
adheres to the expected access policy over the plaintext SBOM.
This is done by comparing the decrypted SBOM tree to the redacted tree using
the embedded sameness proof.
If this sameness proof check passes, the producer countersigns the redacted SBOM prior to distribution,
allowing consumers to gain confidence that the SBOM is a faithful representation of the expected
SBOM information, as generated by the generator, and endorsed by the producer.

%

\subsection{SBOM Distribution}
A producer distributes the redacted SBOM by providing
the distributor a resolvable index value (e.g., a pURL) to allow consumers to discover the SBOM for the artifacts of interest.
In addition to receiving an SBOM, a \system distributor can also verify the signature and counter-signature from the generator and producer, respectively, to ensure that only authenticated SBOMs are available to consumers.

\subsection{SBOM Consumption \& Verification}
Consuming and verifying an in \system SBOM requires a series of verification passes as depicted in Alg.~\ref{alg:consumption}.
First, a consumer must check the authenticity of the $RedactedSBOM$ by verifying both the signature and counter-signature using their public keys $PK_{gen}$ and $PK_{prod}$, respectively (lines 1-6).
After, the AES keys for which the consumer has access with her key $SK_{A}$ are unpacked from the top-level SBOM node (line 7).
Then, the consumer performs a full traversal of the redacted SBOM tree, evaluating each node’s ciphertext policy against its attribute key. Each SBOM node embeds a CP-ABE access tree in its policy field, and the check $satisfies(node.policy)$ determines whether the attributes bound to the consumer’s secret key $SK_{A}$ satisfy this structure. Each node whose policy is satisfied is decrypted using the corresponding AES key, while all others are retained in encrypted form, yielding an SBOM tree containing only the consumer-authorized plaintext (lines 8–16).
Lastly, a consumer can verify a membership proof for a specified decrypted SBOM field $f$ to ensure the generator did not omit information.

\begin{algorithm}
\footnotesize
\caption{Consumption \& Verification}\label{alg:consumption}
\KwData{$RedactedSBOM, SK_{A}, PK_{gen}, PK_{prod}, f$}
\KwResult{$DecryptSBOM$}

    \If{$! VerifySignature(RedactedSBOM.countersig, PK_{prod})$}{
        \Return FAIL\_UNTRUSTED\_SBOM
    }  

    \If{$! VerifySignature(RedactedSBOM.sig, PK_{gen})$}{
        \Return FAIL\_UNTRUSTED\_SBOM
    }
    
    \CommentSty{/* Obtain decryption AES Keys */}
    $AESSet \gets DecryptKeys(SBOMNode, SK_{A})$
    
    \CommentSty{/* Traverse tree and decrypt accessible nodes */}
    $DecryptSBOM \gets new\;SbomTree()$

    \For{$ node \in RedactedSBOM$}{
        \eIf{$satisfies(node.Policy)$}{
        $PlainNode \gets Decrypt(node, AESSet)$
        }
        { 

        $PlainNode \gets node$}
        
        $DecryptSBOM.insert(PlainNode)$
    }

    \CommentSty{/* Compute membership using decrypted tree */}
    $proof \gets computeMembershipProof(DecryptSBOM, f)$
    
    \If{$! verify(proof)$}{
        \Return FAIL\_GENERATOR\_PRODUCER\_LIED
    }

    \CommentSty{/* Return decrypted SBOM tree */}
    
    \Return DecryptSBOM
\end{algorithm}

\Paragraph{Verifier pass.} 
In practice, verifiers may represent trusted parties that have permission to access a whole redacted SBOM.
By appending an \texttt{OR Verifier} with verifier being an access subtree representing the access attributes of a, a trusted third-party, such as an auditor or government agency, can be allowed to decrypt the SBOM and verify its sameness and redistribution congruence integrity properties.


\subsection{Re-distribution \& Composition}

Lastly, a redacted SBOM can be also included inside of \emph{another} SBOM --- a crucial use case for 
the confidential SBOM exchange problem, as artifacts are composed and distributed through the software supply chain.
Perhaps the most notable composition example is that of dependencies being included inside of another artifact.

Diving deeper into the composition step outlined in Algorithm~\ref{alg:redaction},
\system must handle sameness proof generation specially for SBOMs not created by the generator, but rather are provided by the  producer wishing to compose multiple SBOMs.
Specifically, because \system computes sameness proofs upon initial SBOM ingestion, to enable sameness proof verification of a composed SBOM tree $CompSBOM$ without needing the full plaintext tree for an inserted
$RedactedSBOM_i$, 
the generator first recomputes the sameness proof of $CompSBOM$ incorporating the value of the sameness proof at the root of $RedactedSBOM_i$.

Then, the updated sameness proof is embedded when computing the Merkle hash over the remainder of the redacted $CompSBOM$ tree.
This way, the sameness proof can be verified by the producer of the top-level $CompSBOM$, but under the assumption that the inserted $RedactedSBOM_i$ is valid. 






    

%

%

\section{Implementation and Evaluation}
\label{sec:impl-evaluation}
\subsection{Implementation}
We implemented Petra's SBOM parsing and tree construction in Python, while using Rust for CP-ABE functionality and other cryptographic routines.
We used the Maturin framework~\cite{maturin} as well as PyO3~\cite{pyo3} to provide bindings between the two.
Further, we use vendor-provided SBOM libraries for SPDX and CycloneDX translation and support.
In total, the Petra implementation uses less than 1,600 Python LoC and less than 
190 Rust LoC. 

For artifact indexing, we use Package URLs (pURLs)~\cite{purl_spec_2024}, which defines a structured and ecosystem-agnostic format for naming software components (\eg pkg:npm/express@4.17.1). 
This usage is unrelated to “persistent URLs”; our use of pURLs aligns with conventions in both major SBOM standards~\cite{spdx-2.1, noauthor_cyclonedx_nodate}. 

We implemented a KMS for \system by extending the Sigstore project~\cite{newman_sigstore_2022}.
Our modifications allow for sigstore identity proofs (using OIDC) to also pass user attributes A for $SK_A$ generation.
The core set of attributes that are provided by the KMS are usernames, and domains (e.g., foo@bar.com is decomposed as \texttt{user:foo} and \texttt{namespace:bar.com}).
To further provide specific attributes, the KMS can be configured to match particular claims within an OIDC token (e.g., to show control of a particular workload identity, or account-specific attributes).
Lastly, the identity token provided to the KMS is used to derive temporal attributes at a time-window granularity (currently set at one month).

\subsection{Performance Evaluation}
\label{subsec:evaluation}

First, we set off to characterize Petra's practicality by studying the overhead of running Petra.
 We posit that the overhead of using Petra falls within three major categories:
    1) The timing to convert regular SBOMs into SBOM trees,
    2) The storage overhead of a redacted SBOM relative to a regular SBOM, and
    3) The timing for encrypting and verifying properties within a Redacted SBOM.

\Paragraph{Experimental Setup} We carried out the evaluation on a 2.10 GHz Intel Xeon E5620
machine with 256 GB of RAM using a standard Ubuntu 22.04.2 LTS Docker container, a stable
build of Rust (rustc 1.75.0), and Python 3.10.

\Paragraph{Evaluation Policies}
In addition, we aim to understand the effect of specific policies as they apply to a redacted SBOM.
We expect that more complex policies may result in a larger ciphertext, and thus increase the overhead.

 Identifying real-world examples of SBOM access policies was a major challenge.
 To construct representative policies for our evaluation, we analyzed literature and identified two primary categories of sensitive data in SBOMs: proprietary information and weaknesses information~\cite{xia_sbom_study_2023,bi_way_2024,zahan_software_2023,Software_Consumers_Playbook,Software_Suppliers_Playbook,stalnaker_boms_2024}. Then for each type of sensitive data, we map it to claim types in an SBOM (e.g., dependency claims might threaten intellectual property rights, and vulnerability claims might expose security weaknesses). 
 We included security weaknesses, and intellectual property policies in Appendix \ref{appendix: SBOM policies}.


\Paragraph{SBOM Dataset}
We use SBOM datasets from previous literature~\cite{torres_sbom-quality_2023}. After normalization and de-duplication, these datasets contain 3,380 SBOMs from both generated SBOMs (e.g., by a third-party tool), as well as in-the-wild SBOM collections (e.g., identified by indexing platforms such as sourcegraph~\cite{sourcegraph}).This dataset contains SBOMs that cover between 1 and 5002 packages, with an average of 242 per SBOM, a median of 122, and a standard deviation of 408 packages. Our evaluation against this highly variable range of SBOMs reflects the structural complexity and real-world diversity of software artifacts that Petra supports.


\begin{figure}[t]
    \centering
    \includegraphics[width=0.8\linewidth]{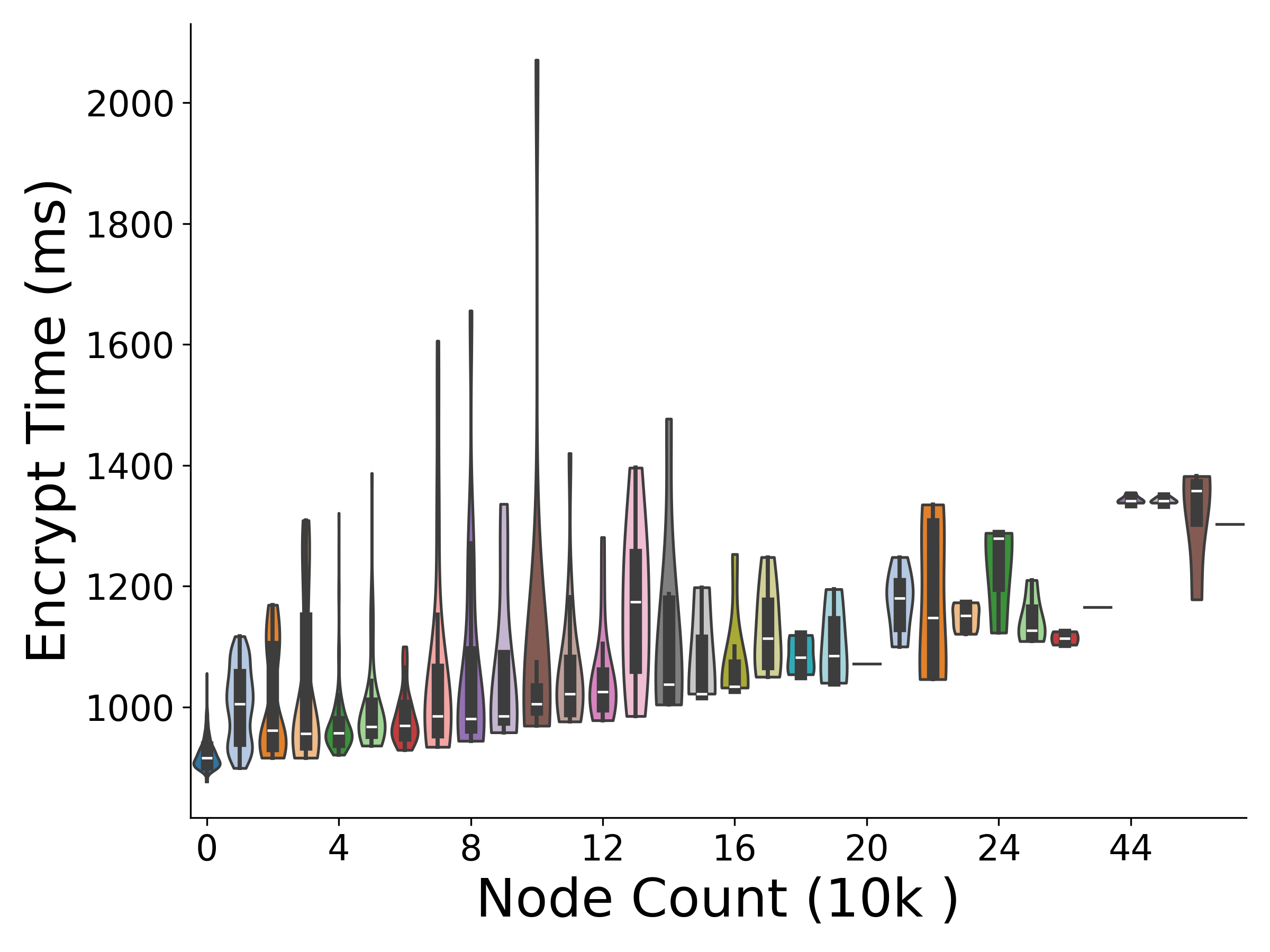}
    \caption{Encryption Overhead for Intellectual Property policy: Decryption overhead matches encryption overhead, with an average deviation of 120.05 ms.}
    \label{fig:encryption-performance}
\end{figure}

\begin{figure}[t]
    \centering
    \includegraphics[width=0.8\linewidth]{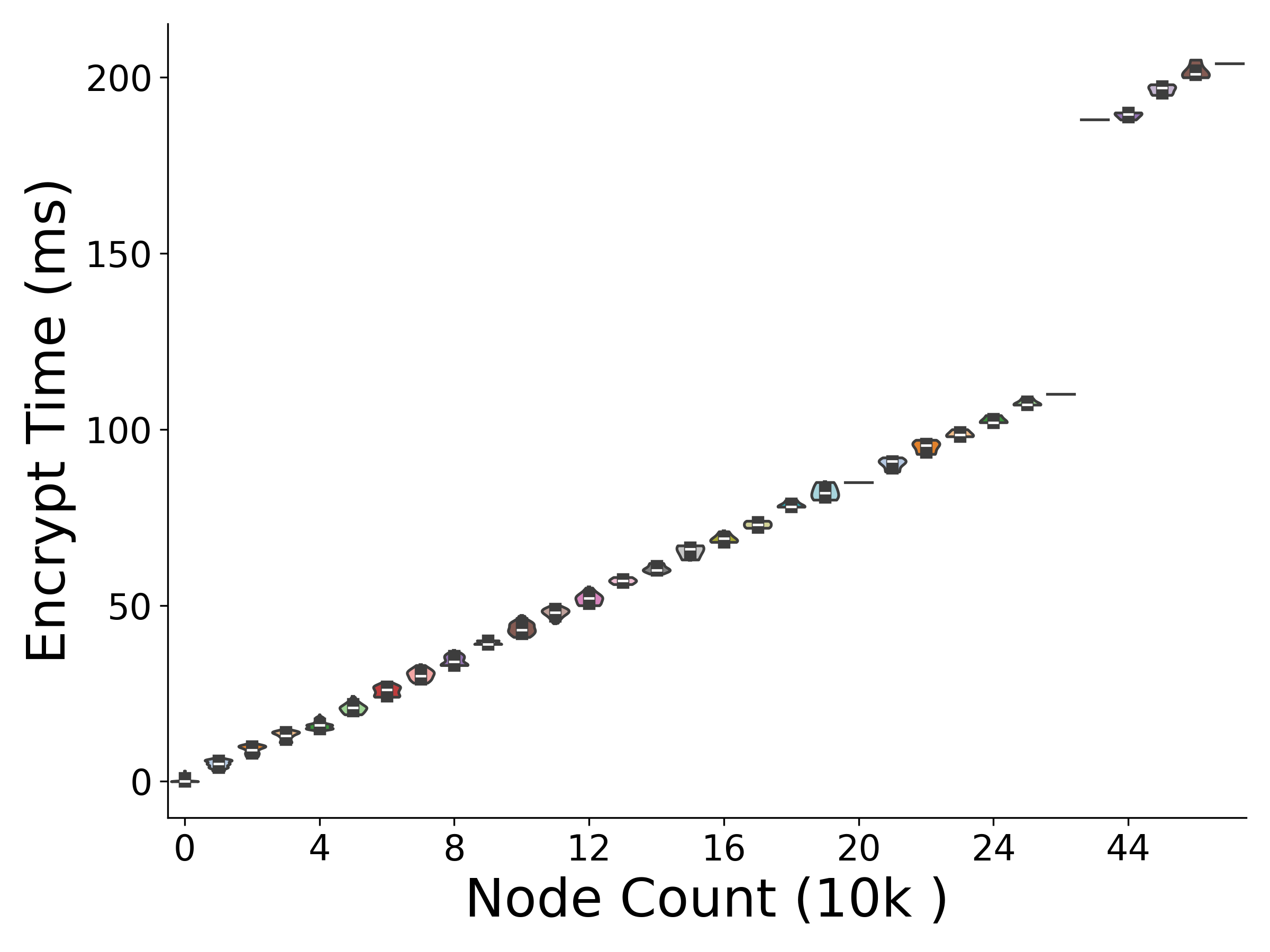}
    \caption{Encryption Overhead for the weaknesses policy: Decryption overhead matches encryption overhead, with an average deviation of 1.16 ms.}
    \label{fig:weaknesses_policy_encryption-performance}
\end{figure}


\subsubsection{\textbf{Storage Overhead}}
We measure the resulting size of Petra trees for the SBOMs in our dataset based on each policy, as well as a non-encrypted SBOM size.
We find that redacted SBOM trees incur less than 13\% increased storage cost on average compared to their unencrypted counterparts.
This cost is dominated by the complexity of the policy.
In the complicated policy case, the cost of encrypting a medium-sized SBOM (i.e., 12MB) is rather stark with more than 13\% increased in the space of the tree, which may make Petra an undesirable system for highly-complicated access scenarios.

However, for our two illustrative policies, a medium-sized SBOM requires only 12\% and 51\% encryption and decryption storage overhead, respectively.
This is acceptable for a system that provides confidential SBOM Exchange according to informal discussion with industry stakeholders.

\subsubsection{\textbf{Performance Overhead}}

We now study the Performance overhead of Petra. 
In this case, we measure the cost incurred by each processing pass (tree, encryption, Merkle-hashing) to provide a granular understanding of each foundational technique. 

Figs.~\ref{fig:encryption-performance} and~\ref{fig:weaknesses_policy_encryption-performance}
illustrate the timing overhead of encrypting an SBOM under the intellectual property and weaknesses policies, respectively. 
We observe a visible discontinuity in the encryption timing results (Fig.~\ref{fig:weaknesses_policy_encryption-performance}) for a small subset of SBOMs. This gap corresponds to a few large SBOMs containing more than 440K nodes. We theorize that these SBOMs exceed CPU-cache capacity, triggering additional memory accesses and a visible discontinuity in the graph. This is an artifact of the "in the wild" dataset we are operating with, where large SBOMs (in terms of tree size) also carry substantial additional data and thus are also larger in storage. 

Unsurprisingly, the decryption cost in our experiments closely mirrors the encryption time, with a mean deviation of 120 ms, indicating symmetry in processing times for encryption/decryption operations. Much like the storage overhead costs, the processing cost is dominated by the CP-ABE pass.

\subsubsection{\textbf{Comparison with Protobom}} To evaluate Petra against existing tooling, we compare it to Protobom~\cite{protobom}, which generates a format-agnostic graph given an input SBOM document. Using the SPDX 2.3 SBOMs in our dataset (due to Protobom’s format support), we measured the number of nodes of their internal graph representations.
We find that Protobom on average produces 93\% fewer nodes than Petra. This reflects the design goals of each system: Protobom prioritizes efficient format translation, while Petra emphasizes preserving compositional relationships. While resulting in a more ``verbose'' tree, our design also facilitates fine-grained access control.

\subsection{Practical Evaluation}
We developed Petra through a research collaboration with an enterprise team managing large-scale SBOMs during an internship at Intel Labs, and its design reflects direct operational and practical security feedback. In enterprise settings, SBOMs are frequently redistributed across organizational and trust boundaries. 

One of the Petra design features requested by the enterprise team is a KMS and verifier as foundational components for practicality: they provide a user-friendly and centrally managed way to issue decryption keys, enforce organizational authorization policies, which relieve producers and consumers from handling cryptographic keys directly. 

However, these services alone are not sufficient. A KMS can issue keys and a verifier can check signatures, but neither enables fine-grained access control, nor do they allow SBOMs to remain verifiable while being redistributed. While an implementation could, in principle, combine decryption with the KMS if fully trusted, our design keeps decryption and verification distributed so that SBOMs can be safely re-distributed and independently verified by arbitrary parties. 

While Petra has not yet been deployed in a production environment, CP-ABE based systems for electronic health records~\cite{etsi2018ts103532} show that such deployments are practical.

\section{Security Analysis}
\label{sec:security-analysis}
Petra's ability to achieve our desired security and confidentiality goals is central.
Here, we study how Petra's design goals in \S~\ref{secsec:design-properties} address threats \textbf{(T1–T7)} in \S\ref{subsec: threat-model}.

At a high-level, we consider two areas of attack in the confidential SBOM exchange threat model.
First, generators or distributors may seek to compromise the integrity of an SBOM to mislead consumers (\S\ref{sec:security-analysis:generator-distributor}). 
Second, consumers may attempt to violate confidentiality by gleaning information from an SBOM without having explicit access (
\S\ref{sec:security-analysis:consumer}).

\subsection{SBOM Integrity Attacks}\label{sec:security-analysis:generator-distributor}

\Paragraph{Tampering with SBOM data (T1):} 
A generator, or a producer that acts as or colludes with the generator, may misbehave by adversarially tampering with SBOM data. 
For instance, the generator might make critical elements inaccessible to consumers by removing a node from the SBOM tree or altering node data to bypass security or compliance audits. Through this, consumers may make a wrong determination about the security posture of a software artifact. 

\system's sameness property \textbf{(I1)} addresses this threat by making structural changes between plaintext and redacted SBOMs detectable by consumers using Merkle trees and membership proofs.
That said, consumers cannot detect tampering if a leaf (and any sibling) nodes are removed for SBOM fields that a consumer cannot decrypt, because this scenario looks indistinguishable from an SBOM with a stricter access policy. 

\Paragraph{Split view attacks (T2):}
A distributor may collude with a generator to show different consumers two different SBOMs for the same artifact. In this attack, the attacker tries to fool a victim consumer $V$ into using artifact $A$ by sending her a compromised SBOM $S_V$ (e.g., omitting a vulnerability). 
At the same time, they send another consumer $C$ the faithful SBOM $S_A$, allowing her to detect the vulnerability. 

\system provides non-equivocation \textbf{(I3)} by giving $V$ and $C$ a means to observe each other's SBOMs during SBOM redistribution of the faithful $S_A$ via another uncompromised SBOM, revealing this attack.
That is, because the redistributed $S_A$ and $S_V$ would have different Merkle root hashes, even though they represent the same artifact $A$, the victim consumer $V$ would detect the attack. 
In addition, the signature/counter signature scheme, provides non-repudiation for generators and producers, which allows consumers to verify they are both looking at an SBOM provided by the same party.
Lastly, signed Merkle tree roots can optionally be published to a public transparency log to support global visibility and facilitate split view detection.

\Paragraph{Corrupting SBOMs redistributed to downstream consumers (T3):} A distributor, or compromised consumer during redistribution, may tamper with an SBOM while in transit. 
This causes a denial-of-service (DoS), rendering the SBOM unusable to downstream consumers. 
However, the redistribution congruence property \textbf{(I2)} of SBOM trees addresses this threat because this tampering would be detectable by any consumer with sufficient access to decrypt the SBOM and validate its contents or structure.


    

%

\subsection{SBOM Confidentiality Attacks}\label{sec:security-analysis:consumer}

\Paragraph{Unauthorized SBOM data leaks (T4):} 
A consumer with access to a plaintext SBOM may misuse or maliciously share sensitive information contained therein to identify
points of access to a system or vulnerability information for competitive analysis or more a harmful software producer compromise~\cite{dalia_sbom_2024, kloeg_charting_2024, xia_sbom_study_2023, zahan_software_2023}.
Thus, \system provides semantic confidentiality \textbf{(C1)} by enabling software producers to redact SBOMs using selective encryption, limiting the risk of unauthorized SBOM data leaks in transit and at rest.

\Paragraph{Dictionary attacks (T5):} An attacker could attempt to carry out a dictionary attack on specific nodes.
This can happen by using the sameness proof hashes, and the fact that it is likely some fields have a very limited ``key space''.
License elements, for example, have a limited number of field values --- 677 on the latest SPDX list~\cite{spdx-license-list-xml} --- which could be enumerated by an attacker.

However, \system preserves semantic confidentiality \textbf{(C1)} through sameness proofs that use a random salt that is hashed and encrypted along with the node data, requiring an attacker to not only enumerate possible values, but also the nonce key space (256 bits of key space in the current Petra implementation), which makes such attacks computationally infeasible.

\Paragraph{Collusion to decrypt inaccessible information (T6):} 
A group of consumers with non-overlapping attributes may try to collude to access more information than what they can individually access. 
However, Petra's use of CP-ABE for redaction offers collusion resistance \textbf{(C2)}, which addresses this threat because the scheme uses single decryption
keys for data accessible to holders of specific $N:N$ combinations of attributes (\ie permissions) to resist collusion between parties that have access to only a subset of the needed permissions.

\Paragraph{Inferring inaccessible SBOM data (T7):} 
A consumer may attempt to infer redacted SBOM data using partial access to the same SBOM or through other related SBOMs. Petra addresses two classes of information inference attacks.

\textbf{Intra-SBOM Inference Attacks:} 
An attacker with access to a portion of an SBOM using their attributes tries to infer other aspects of the redacted SBOM.

\noindent Of these, we identify two variants:

\begin{itemize}
    \item \textbf{Structural Inference}: An attacker seeks to infer the \emph{field name} in encrypted nodes.
    \item \textbf{Semantic Inference}: An attacker seeks to infer the \emph{field name and value} in encrypted nodes.
\end{itemize}

An attacker may, for example, study the padding properties of field names and values.
This risk can be reduced in practice by using a reasonable padding value at a slight cost on storage.

Further, an attacker can attempt to infer information by looking at the structure of the redacted SBOM tree.
Certain aspects of an SBOM will be biased to some tree structures;
for example, a dependency list is often represented by a complex node with a large number of children.
As such, it is possible an attacker learns certain parts of the tree are likely particular complex fields.

\system preserves semantic confidentiality \textbf{(C1)} in this attack scenario because SBOM tree leaf nodes, while carrying the most crucial information, are unlikely to leak any information from their structure because individual field nodes are encrypted in a uniform way using a non-deterministic encryption scheme. 


\textbf{Inter-SBOM Inference Attacks:} An attacker with access to an SBOM may use it to try to infer information about another SBOM they do not have access to. 
In this scenario, the attacker has the key to decrypt certain SBOM fields in $SBOM_A$ and tries to use this information to infer the value of those fields in an $SBOM_B$. 
However, \system preserves semantic confidentiality \textbf{(C1)} by reducing the probability of an attacker inferring a field \emph{value}, even if they were equal in the two SBOMs.
That is, by employing encryption schemes that are non-deterministic and dependent on the input access policy, an attacker is highly unlikely to succeed in correlating nodes between two SBOMs.

\section{Discussion}

The Petra construction outlined above helps further both SBOM confidentiality and transparency.
We discuss the following limitations and future extensions.

\Paragraph{SBOM quality \& accuracy.}
Petra’s cryptographic guarantees and access control do not depend 
on the quality and completeness of underlying SBOM data. Nevertheless, improvements in SBOM quality and accuracy could enable \system to enforce richer SBOM redaction policies.

\Paragraph{Expanding to other types of metadata.} Currently, Petra focuses mainly on SBOMs, and a promising future direction is adding support for other supply chain metadata formats such as VEX, OpenVEX, or SLSA.  
By expanding Petra to handle these other formats, the system could offer a more comprehensive solution to secure and audit software supply chains.  
However, integrating these formats would introduce challenges related to the diversity in their structure, semantics, and encryption needs. 
Future work could explore efficient ways to harmonize these differences, ensuring Petra remains both flexible, semantically consistent, and scalable. 


\Paragraph{Software update history.} A desirable property of SBOM exchange is capturing the over-time evolution of software in SBOMs~\cite{xia_sbom_study_2023}.
Though at first sight the composition system within Petra may allow to \emph{append} new information in redacted SBOMs (e.g., new vulnerability information), this is not sufficient.
Instead, a mechanism that allows for appending new information, but also removing outdated information, is necessary.
For this, other backing data structures such as a $Merkle^2$~\cite{popa2021_merkle2} tree may allow for both additions, erasures and replacements on an SBOM tree.

\Paragraph{Policy updates post-publication.}
 Petra’s design enables partial updates and avoids full SBOM regeneration in response to policy update. Since each node in the SBOM tree is individually encrypted and tied to its specific access policy, only nodes whose access policies have changed need to be re-encrypted. Corresponding Merkle proofs are updated for those nodes and any intermediate nodes affected by the change. 
 Importantly, redistribution does not need to be repeated if policy changes are made to individual fields, updating a policy and re-encrypting a single field does not invalidate redistributed copies. 
 This design allows producers to modify or refine access control over time(e.g., granting broader visibility to a previously restricted field) without invalidating already-redistributed SBOMs. However, redistribution becomes necessary when structural changes occur.
 
\Paragraph{Semantic-aware policies.} \system currently supports SBOM redaction policies at the granularity of fields and complex SBOM elements, and treats the SBOM as a non-semantic tree (i.e., a collection of cryptographically protected claims rather than a semantically interpreted document). As a result, a redactor with a weak policy could encrypt data in a way that undermines SBOM utility without violating any cryptographic guarantees.
For example, if a redactor encrypts all related fields with a different policy, no consumer can make any meaningful verification. 

This limitation reflects a deliberate separation between cryptographic verification and semantic interpretation.
\system supports this separation by allowing producers and designated verifiers to access full-SBOM plain-text to validate complete sameness, while consumers receive structurally verifiable views limited to the information they are authorized to access. 
A malicious producer cannot undetectably alter or remove information within the verifiable portion of an SBOM, as any modification would invalidate the corresponding Merkle proofs or expose a redaction marker, assuming \system uses a collision-resistant hash function. However, determining whether redaction conceals semantically relevant information is inherently a policy-level concern.
Thus, a semantic-aware SBOM translation algorithm could further limit the possibility of undermining SBOM utility and should be trivial to implement on top of the Petra SBOM code. 

\Paragraph{Consumer policies.} A potential future extension of the current construction is allowing consumers to define security policies for consumed SBOMs.
A separate role such as a policy issuer could obtain policy requests from consumers and enforce their usage at redaction.
For example, by using Homomorphic-Policy Attribute-Based Encapsulation~\cite{chotard2016homomorphic}, a consumer may further append attributes to the SBOM tree node policy before redistribution to limit access to sub-elements on SBOMs under their discretion. 
However, this feature may further introduce complexity and risk practical deployment.

\Paragraph{Further confidentiality properties.} There are a number of additional confidentiality properties that may be of interest in future extensions of \system, such as node compression or encrypting node references to hide the structure of an SBOM.
However, we argue that the current design allows for easier membership proof, and a simpler composition flow.

\section{Conclusion}
\label{sec:conclusion}

In this paper, we presented the first formally-defined abstract SBOM exchange model.
From it, we extended this model to consider confidentiality expectations within the involved parties.
Ultimately, these two notions present a way to develop new systems and protocols to address confidentiality and integrity expectations of software users.

Beyond the modelling effort, we presented Petra, the first system to provide confidential SBOM Exchange capabilities, while maintaining tool compatibility expectations.
While Petra is not the only design that can achieve these goals, it satisfies the majority of design expectations for a system that provides SBOM Exchange.
Due to its design, Petra can be integrated with other SBOM distribution efforts within industry and open source.
We are currently engaging with open source projects to integrate \system structures and processes to achieve stronger confidentiality properties.

\section*{Acknowledgment}
\label{sec:ack}

We thank anonymous reviewers for their constructive feedback. We also thank Intel Corporation for supporting this research through a single-PI grant. 
\appendix
\input

\section{Ethical Considerations}
Petra is primarily concerned with confidentiality in the exchange of SBOM data.
We do not expect any ethical harms from providing software producers with better control over their data. 
Instead, Petra supports a goal that aligns with widely accepted values of security and transparency, namely enabling organizations to protect confidential details (e.g., intellectual property and vulnerability information) while still providing accountability through auditable SBOM exchange.

To help ensure that our design meets real-world needs and expectations, we engaged with both open-source and enterprise stakeholders during system development. 
Their feedback helped shape the design of our redaction and verification mechanisms, and practical adoption concerns. 
We believe Petra provides a balanced solution that protects confidentiality, enables trustworthiness, and avoids ethical pitfalls that could arise from restricting information access arbitrarily.

\Paragraph{Affected stakeholders and implications: } Petra impacts several stakeholder groups.

\Paragraph{Software producers (open source and enterprise vendors)} benefit from the ability to share SBOMs without disclosing sensitive information, addressing a core barrier to practical SBOM adoption. Petra’s cryptographic sameness proofs limit a producer’s ability to misrepresent SBOM contents without detection by authorized verifiers. However, Petra does not prevent producers from choosing overly restrictive redaction policies; determining whether a policy is appropriate remains a governance and compliance responsibility.

\Paragraph{Software consumers}
including those that subsequently act as producers by redistributing composed artifacts, gain stronger assurances that redacted SBOMs faithfully correspond to the plaintext SBOM, improving confidence in analyses like SBOM-based risk assessments. At the same time, consumers may receive only partial visibility depending on access policies, which may constrain certain analyses.~\system makes these limitations explicit and verifiable, rather than implicit.

\Paragraph{Verifiers (auditors, regulators, and compliance authorities)}
are privileged stakeholders that \system enables to verify SBOM integrity without requiring universal disclosure. This supports regulatory oversight while minimizing unnecessary exposure of sensitive data. Petra does not replace regulatory judgment or audit policy; it provides cryptographic mechanisms that they may choose to rely on.

\Paragraph{Other SBOM exchange roles (generators, distributors, KMS)}
are affected differently by \system’s design. Distributors benefit from reduced exposure to confidential SBOM data, as they are expected to store only redacted SBOMs and don't get decryption keys for redacted fields. Generators, by contrast, operate on software artifacts and plaintext SBOMs in order to perform SBOM generation, redaction, and adding integrity proofs, and therefore remain trusted with access to sensitive information. Key management services (KMS) issue and manage cryptographic keys and enforce access control via attributes, but do not require access to SBOM contents.

\system reduces the need for distributing plaintext SBOMs by restricting plain-text SBOM access to authorized roles, but it assumes correct cryptographic implementation and key management practices; operational failures in these components remain outside the scope of Petra’s ethical guarantees.

\section{Open Science}
\system artifacts are open source including source code and evaluation, under Apache License 2.0 and are available at https://doi.org/10.5281/zenodo.17906622 . The bom-shelter dataset we used is available at https://doi.org/10.5281/zenodo.17859760 .
\bibliographystyle{plain}
\bibliography{main}

\appendix
\section*{Appendix}
\label{appendix}
\begin{figure*}
    \includegraphics[width=\textwidth]{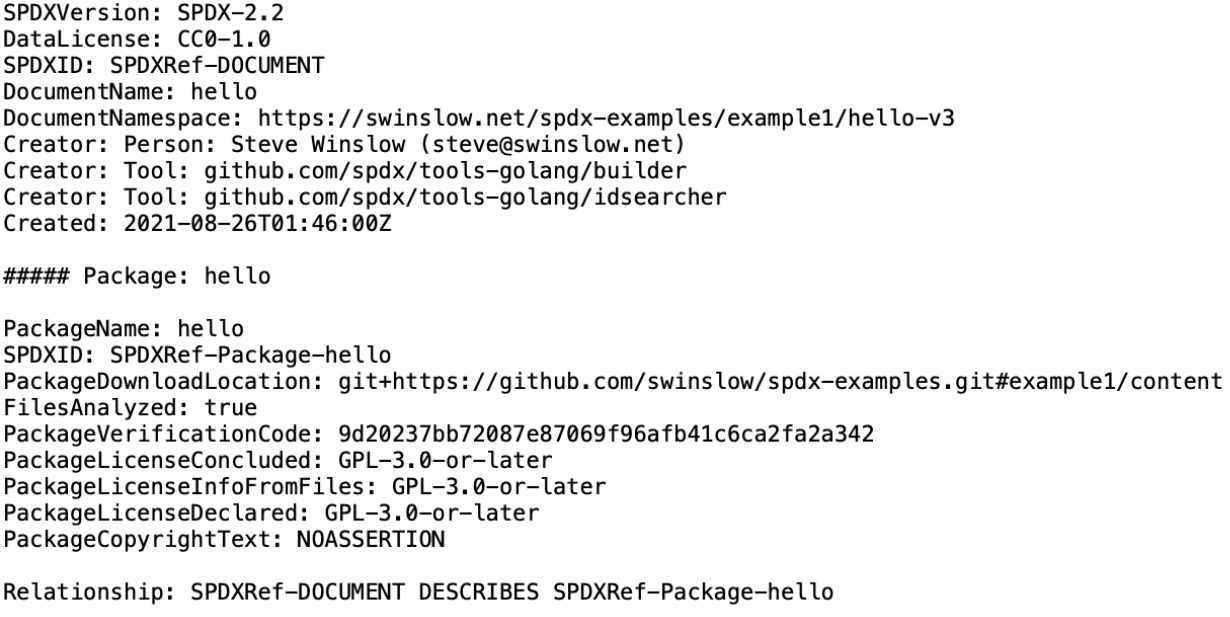}
    \caption{An example SPDX document describing the package \texttt{hello}. It includes metadata such as licensing information, package verification codes, file checksums, and relationships between files.}

 \label{fig:sbom-example}
\end{figure*}
\section{SBOM Example}
\label{appendix: SBOM example}
Figure \ref{fig:sbom-example} represents SBOM in SPDX format for the \texttt{hello} package that contains the source file \texttt{hello.c}, a \texttt{Makefile} for building it, and the compiled binary \texttt{hello}. All files are licensed under \texttt{GPL-3.0-or-later}. The SPDX document provides checksums for verification and describes relationships, showing that the binary is generated from the source code using the \texttt{Makefile}.

\section{SBOM policies}
\label{appendix: SBOM policies}
\begin{figure*}
    \includegraphics[width=\textwidth]{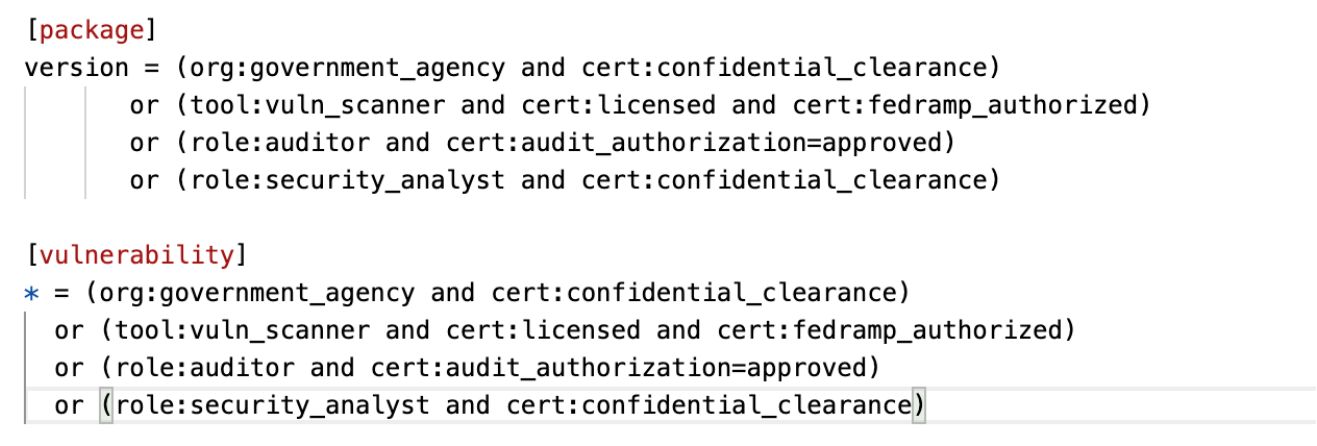}
    \caption{Policy to restrict the access to information that might reveal weaknesses of the software artifact that the SBOM describes.}

    \label{fig:weaknesses-policy}
\end{figure*}
\begin{figure*}
    \includegraphics[width=\textwidth]{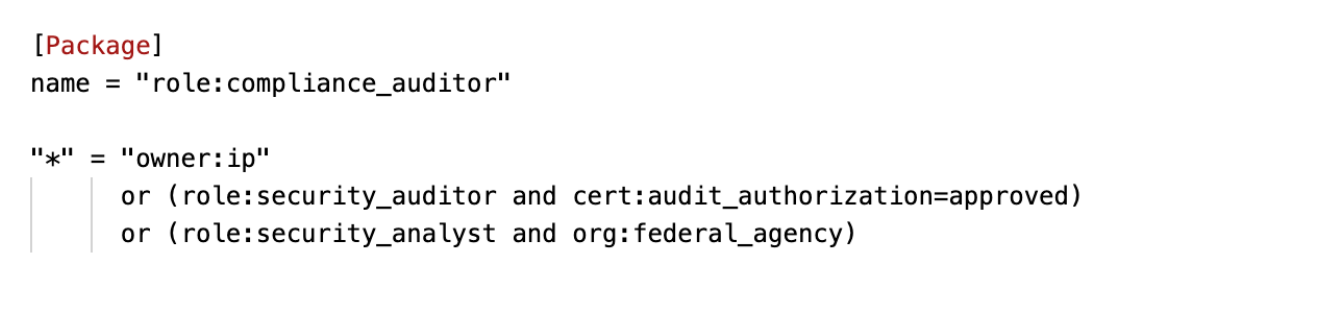}
    \caption{Policy to restrict the access to proprietary information of the software artifact that the SBOM describes.}

    \label{fig:ip-policy}
\end{figure*}
Figure~\ref{fig:weaknesses-policy} represents Petra policy used to control access to information that can reveal weaknesses, and Figure~\ref{fig:ip-policy} represents Petra policy used to control access to proprietary information of the software artifact that an SBOM describes
.

\end{document}